\documentclass[preprint,aps,amsmath,showpacs,showkeys,nofootinbib,
superscriptaddress]{revtex4}
\usepackage[dvips]{graphicx}

\begin{document}

\title{\bf Multi-$\bar K$ nuclei and kaon condensation}

\author{D.~Gazda}
\email{gazda@ujf.cas.cz}
\affiliation{Nuclear Physics Institute, 25068 \v{R}e\v{z}, Czech Republic}

\author{E.~Friedman}
\email{elifried@vms.huji.ac.il}
\affiliation{Racah Institute of Physics, The Hebrew University,
Jerusalem 91904, Israel}

\author{A.~Gal}
\email{avragal@vms.huji.ac.il}
\affiliation{Racah Institute of Physics, The Hebrew University,
Jerusalem 91904, Israel} 

\author{J.~Mare\v{s}}
\email{mares@ujf.cas.cz}
\affiliation{Nuclear Physics Institute, 25068 \v{R}e\v{z}, Czech Republic}

\date{\today} 

\begin{abstract}
We extend previous relativistic mean-field (RMF) calculations of 
multi-$\bar K$ nuclei, using vector boson fields with SU(3) PPV 
coupling constants and scalar boson fields constrained phenomenologically. 
For a given core nucleus, the resulting $\bar K$ separation energy 
$B_{\bar K}$, as well as the associated nuclear and $\bar K$-meson 
densities, saturate with the number $\kappa$ of $\bar K$ mesons for 
$\kappa > \kappa_{\rm sat} \sim 10$. Saturation appears robust against 
a wide range of variations, including the RMF nuclear model used and the 
type of boson fields mediating the strong interactions. Because $B_{\bar K}$ 
generally does not exceed 200~MeV, it is argued that multi-$\bar K$ nuclei 
do not compete with multihyperonic nuclei in providing the ground state 
of strange hadronic configurations, and that kaon condensation is unlikely 
to occur in strong-interaction self-bound strange hadronic matter. 
Last, we explore possibly self-bound strange systems made of 
neutrons and ${\bar K}^0$ mesons, or protons and $K^-$ mesons, and study 
their properties.  
 
\end{abstract}



\maketitle 

\newpage

\section{Introduction and Overview} 
\label{sec:intro}

Kaon condensation in dense matter was proposed over 20 years ago by 
Kaplan and Nelson \cite{KNe86,NKa87}. It is necessary to distinguish 
in this context between $K$ mesons and $\bar K$ mesons which interact 
quite differently with matter. The empirical evidence from $K^-$ atoms 
is that the $\bar K$-nuclear interaction is strongly attractive, and 
absorptive as well, with typical values of $150-200$~MeV attraction
at nuclear-matter density $\rho_0$, as reviewed recently by Friedman 
and Gal \cite{FGa07}. A strong nuclear attraction of somewhat less than 
100 MeV at $\rho_0$ for $K^-$ mesons, compared to a weak repulsion of 
order 25 MeV for $K^+$ mesons, follows from observations of enhanced 
near-threshold production of $K^-$ mesons in proton-nucleus collisions 
at GSI \cite{SBD06}. This weakly repulsive nature of the $K^+$-nuclear 
interactions was quantified a long time ago, starting with Dover and 
Moffa \cite{DMo77}, and is also reviewed in Ref.~\cite{FGa07}. 
Given the distinction between the nuclear interactions of $K$ mesons 
and $\bar K$ mesons, the term {\it kaon condensation} is used loosely here 
and elsewhere to mean $\bar K$ condensation.

Neutron stars, with a density range extending over several times 
$\rho_0$, offer the most natural dense systems where kaon condensation 
could be realized; see Refs.~\cite{Lee96,PBP97,HHJ00,HPa00,Gle01,RSW01} 
for comprehensive reviews of past work. We note that in {\it Heaven}, 
for neutron stars, weak-interaction time scales of order $10^{-8}$~s 
and longer are operative, enabling strangeness-changing processes such 
as $e^- \to K^- + \nu_e$ to transform high-pressure dense electrons to 
$K^-$ mesons once the effective mass of $K^-$ mesons drops down below 
approximately 200 MeV. Under some optimal conditions, recalling that 
$\bar K$ mesons undergo attraction of order 100 MeV per density unit of 
$\rho_0$ \cite{SMi96}, kaon condensation could occur at densities about 
$3\rho_0$, depending on the way hyperons enter the constituency of neutron 
stars as first recognised by Ellis, Knorren and Prakash \cite{EKP95}. 
However, on {\it Earth} under laboratory conditions, 
strong-interaction time scales of order $10^{-23}$~s are operative; 
processes of equilibration and hadronization subsequent to 
dense-matter formation in heavy-ion collisions occur over much shorter 
times than those controlling the composition of neutron stars. 
If antikaons bind strongly to nuclei, according to a scenario spelled 
out recently by Yamazaki {\it et al.}~\cite{YDA04}, then $\bar{K}$ mesons 
might provide the relevant physical degrees of freedom for self-bound 
strange hadronic matter that would then be realized as multi-$\bar{K}$ 
nuclei. It requires that the $\bar K$ separation energy $B_{\bar{K}}$ 
beyond some threshold value of strangeness exceeds $m_Kc^2 + \mu_N 
- m_{\Lambda}c^2 \gtrsim 320$~MeV, where $\mu_N$ is the nucleon chemical 
potential, thus allowing for the conversion $\Lambda \to \bar{K} + N$ 
in matter. For this strong binding, $\Lambda$ and $\Xi$ hyperons would 
no longer combine macroscopically with nucleons to compose the more 
conventional kaon-free form of strange hadronic matter \cite{SBG00}. 
$\bar K$ mesons will then condense macroscopically. However, the nuclear 
densities encountered in these strange hadronic nuclei are somewhat less 
than the typical $3\rho_0$ threshold required to lower sufficiently the 
$\bar{K}$ energy in matter to reach condensation. Yet, precursor phenomena 
to kaon condensation in nuclear matter could occur at lower densities as 
soon as $B_{\bar{K}}$ exceeds the combination 
$m_Kc^2 + \mu_N - m_{\Sigma}c^2 \gtrsim 240$~MeV. In this case, the only 
mechanism underlying the widths of multi-$\bar{K}$ states is the fairly weak 
conversion $\bar K NN \to \Lambda N$.   

Recently we have reported on preliminary calculations of multi-$\bar{K}$ nuclear
configurations \cite{GFG07} using the relativistic mean-field (RMF) methodology,
constrained by $\bar K$-nucleus phenomenology. It was found that the nuclear
and $\bar K$ densities behave regularly on increasing the number of antikaons
embedded in the nuclear medium, without any indication for abrupt or substantial
increase, and that the $\bar K$ separation energy saturates. Roughly speaking,
the heavier the nucleus is, the more antikaons it takes to saturate the
separation energies, but even for $^{208}{\rm Pb}$ the number required does not
exceed approximately 10. Because the calculated $\bar K$ separation energies
$B_{\bar K}$ do not generally exceed 200 MeV, for input binding in the accepted
``deep-binding'' range $B_{\bar K} \sim 100-150$~MeV for a single $\bar K$ meson
\cite{MFG05,MFG06,Wei07}, it was deemed unlikely that kaon condensation occurs
in nuclear matter. This leaves antikaons in multi-${\bar K}$ nuclei comfortably
above the range of energies appropriate to (hyperonic) strange hadronic matter
\cite{SBG00}. In the present article we discuss the full scope of these
calculations demonstrating the robustness of this saturation property. In
particular we study the sensitivity of the results to the nuclear equation of
state used, through the nonlinear RMF version employed, and the role of ``hidden
strangeness'' isoscalar meson fields beyond the standard isoscalar, scalar
($\sigma$), and vector ($\omega$) meson fields. Although both $\sigma$- and
$\omega$-meson fields mediate attraction between ${\bar K}$ mesons and nucleons,
they play different roles for the interactions within ${\bar K}$ mesons,
similarly to the pattern well known for nucleons. The $\sigma$ meson induces
attraction, whereas the $\omega$ meson induces repulsion. If the ${\bar K}$-meson
couplings were exclusively limited to scalar-meson fields, the resulting ${\bar
K}$-meson separation energies would not have saturated. However, chiral model
studies of ${\bar K} N$ low-energy phenomenology give a clear evidence in favor
of the lowest-order Tomozawa-Weinberg {\it vector} interaction, which in terms
of meson exchanges is equivalent to vector-meson exchanges with purely F-type
SU(3) pseudoscalar-pseudoscalar-vector (PPV) vertices \cite{Wei07}. Our
philosophy in this work is to use these vector-meson fields coupling constants
as they are, augmenting the ${\bar K}$-nucleus vector interaction by additional {\it
scalar} couplings such that $B_{\bar K} \sim 100-150$~MeV holds for
single-${\bar K}$ nuclei.  We find no precursor behavior to kaon condensation
for $\bar K$ mesons in self-bound nuclear matter. 

We also explore in this work exotic strange self-bound configurations 
where ${\bar K}$ mesons are bound to either neutrons or protons.  
The simple example of a quasibound $K^-pp$ system (and thus also its 
charge-symmetric partner ${\bar K}^0 nn$) recently calculated solving 
Faddeev coupled-channel equations \cite{SGM07a,SGM07b,ISa07}, clearly 
demonstrates that ${\bar K}$ mesons can bind together nuclear clusters 
that are otherwise unbound. The point here is that the underlying $K^-p$ 
and ${\bar K}^0 n$ interactions (each with equally mixed $I=0$ and $I=1$ 
components) provide considerably more attraction than the purely $I=1$ 
$K^-n$ and ${\bar K}^0 p$ interactions. The RMF calculations reported here 
start with eight neutrons, showing that a finite number of neutrons can be made 
self bound by adding together a few ${\bar K}^0$ mesons, with ${\bar K}$ 
separation energies of order $B_{\bar K} \sim 50-100$~MeV. We study the role 
of the isovector $\rho$ meson in stabilizing these exotic configurations,
owing to its role in distinguishing between the underlying $I=0$ and $I=1$ 
$\bar K N$ interactions. We find that the emergent stable neutron 
configurations are more tightly bound than in the corresponding ordinary
nuclei with $N \approx Z$ along the stability valley, and the neutron 
single-particle spectra display substantial rearrangement.  
However, these exotic configurations are found to be unstable against 
charge-exchange ${\bar K}^0~+~n~\to~K^-~+~p$ reactions. 

In Sec.~\ref{sec:method} we briefly outline the RMF methodology and discuss 
the ${\bar K}$ coupling constants to the meson fields used in the present 
calculations. Results are shown and discussed in Sec.~\ref{sec:res} for 
${\bar K}$ separation energies and density distributions, also displaying the 
dependence on the type of nonlinear RMF model used and the contribution of 
specific meson fields to the energy systematics and particularly to maintaining 
saturation in a robust way. A separate subsection is devoted to the study 
of exotic multi-${\bar K}$ ``nuclei'' with neutrons only.
Again, binding energies and densities are discussed, plus rearrangement 
features of the neutrons in the ${\bar K}$-extended mean field. 
We conclude with a brief summary in Sec.~\ref{sec:concl}.

\section{Methodology}
\label{sec:method}

\subsection{RMF equations of motion}

Bound nuclear systems of nucleons and several $\bar{K}$ mesons are
treated in this work within the RMF framework,
where the interactions among hadrons are mediated by the exchange
of scalar- and vector-meson fields. The model
Lagrangian consists of a standard nuclear part ${\cal L}_N$
and the Lagrangian density ${\cal L}_K$ describing the kaonic
sector:
\begin{equation}
{\mathcal L}_{K} = \left( {\mathcal D}_\mu K \right)^\dagger
\left( {\mathcal D}^{\,\mu} K \right) -m_K^2 K^\dagger K
+g_{\sigma K} m_K K^\dagger K \, \sigma
+g_{\sigma^* K} m_K K^\dagger K \, \sigma^*   \:.
\label{eq:lagrangian} 
\end{equation}
Here,
\begin{equation}
{\mathcal D}_\mu \equiv
\partial_\mu
+ {\rm i}\, g_{\omega K}\, \omega_\mu + {\rm i}\, g_{\rho K}\,
\vec{\tau} \cdot \vec{\rho}_\mu + {\rm i}\, g_{\phi K}\, \phi_\mu
+ {\rm i}\, e\, \textstyle\frac{1}{2}\displaystyle(1+\tau_3) A_\mu
\:,
\label{eq:Del}
\end{equation}
and $K$ ($K^\dagger$) denotes the kaon (antikaon) doublet. To be specific, 
we discuss $K^-$-nuclear systems. Similar expressions hold for ${\bar K}^0$ 
mesons. In addition to a scalar-meson field $\sigma$ and to vector-meson 
fields $\omega$ and $\rho$ normally used in purely nuclear RMF calculations, 
we also considered meson fields that couple exclusively to strangeness 
degrees of freedom, a scalar $\sigma^*$, and a vector $\phi$. 
Standard techniques yield a coupled system of equations of motion for 
nucleons and all meson mean fields involved; we refer the reader to our 
earlier work~\cite{GFG07} for details. Here it suffices to recall that 
the presence of $\bar{K}$ meson(s) induces additional source terms in 
the Klein-Gordon (KG) equations for the meson (mean) fields. In the case of 
$K^-$ mesons, the source terms contain the $K^-$ density 
\begin{equation}
\label{eq:kdens}
\rho_{K^-}=2
(E_{K^-}+g_{\omega K}\,\omega_0+g_{\rho K}\,\rho_0+g_{\phi K}\,\phi_0+e\, A_0)
K^-K^+ \, ,
\qquad \int {\rm d}^3 x\, \rho_{K^-} = \kappa
\:, 
\end{equation}
where $E_{K^-}={\rm i}\,\partial_t K^-$. Hence, the $\bar K$ mesons 
modify the scalar and vector potentials that enter the Dirac equation 
for nucleons, thus leading to rearrangement of the nuclear core. The 
polarized nucleons, in turn, modify the $\bar K$-nucleus interaction. This 
calls for a self-consistent procedure for solving the equations of motion.

In our model, the KG equation of motion for the $K^-$ meson
acquires the form 
\begin{equation}
\label{eq:Kkg}
[-\nabla^2-E_{K^-}^2 +m_K^2 + {\rm Re}\,\Pi_{K^-} \,]K^-=0 \:,
\end{equation} 
with the $K^-$ self-energy given by 
\begin{eqnarray}
\label{eq:self} 
{\rm Re}\,\Pi_{K^-}=
&-&(g_{\sigma K}m_K\sigma_0 + g_{\sigma^* K}m_K\sigma^{*}_0)
-2E_{K^-}(g_{\omega K}\omega_0+g_{\rho K}\rho_0+g_{\phi K}\phi_0+eA_0)
\\ \nonumber
&-&(g_{\omega K}\omega_0+g_{\rho K}\rho_0+g_{\phi K}\phi_0+eA_0)^2 \:.
\end{eqnarray}
Of the three terms on the right-hand side (rhs) of Eq.~(\ref{eq:self}), 
the first one is a scalar-meson contribution, whereas the other two 
terms are vector-meson contributions. The scalar contribution is sometimes 
lumped together with the kaon mass $m_K$ to form a density-dependent 
effective kaon mass $m^*_K$ via 
\begin{equation} 
\label{eq:meff} 
{m^*_K}^2~=~{m_K}^2~-~g_{\sigma K}m_K\sigma_0~-~g_{\sigma^* K}m_K\sigma^*_0\:. 
\end{equation}
Finally, to account for $K^-$ absorption in the nuclear medium, the 
self-energy $\Pi_{K^-}=2E_{K^-}V^{K^-}_{\rm RMF}$ in Eq.~(\ref{eq:Kkg}) was 
made complex by adding ${\rm Im}\,\Pi_{K^-}$ and the real energy $E_{K^-}$ 
was replaced by $E_{K^-}-{\rm i}\Gamma_{K^-}/2$. The imaginary part of the 
self-energy, ${\rm Im}\,\Pi_{K^-}$, was taken from optical model 
phenomenology, with a strength fitted to $K^-$ atomic data \cite{FGM99} 
and with energy dependence that follows the reduced phase space for the 
decaying initial state. We assumed two-body final-state kinematics for 
the decay products in the absorption channels 
${\bar K}N \rightarrow \pi Y$ ($Y=\Sigma,\; \Lambda$) (80\%) and 
${\bar K}NN\rightarrow YN$ (20\%) with branching ratios indicated in 
parentheses \cite{MFG05,MFG06}. 

The set of coupled equations containing the Dirac equation for nucleons,
the KG equations  for the meson mean fields and for antikaons was solved 
fully self-consistently using an iterative procedure. This appeared crucial 
for the proper evaluation of the dynamical effects in nuclei with 
$\kappa~(\kappa=1,2,3,\cdots)~{\bar K}$ mesons. The ${\bar K}$ separation 
energy 
\begin{equation} 
\label{eq:sep}
B_{\bar K}=B[A,Z,\kappa {\bar K}]-B[A,Z,(\kappa-1){\bar K}] \:, 
\end{equation} 
where $B(A,Z,\kappa {\bar K})$ is the binding energy of the 
$\kappa {\bar K}$-nuclear system, contains mean-field contributions 
due to rearrangement of the nuclear core.

\subsection{Choice of parameters}

For the nucleonic Lagrangian density ${\mathcal L}_N$ we used the RMF 
parameter sets NL-SH \cite{SNR93} and NL-TM1(2) \cite{STo94} which have 
been successfully used in numerous calculations of various nuclear systems. 
For the (anti-)kaon coupling constants to the vector-meson fields, we used 
a purely F-type SU(3) symmetry, $\alpha_{\rm V} \equiv F/(F+D) = 1$:  
\begin{equation} 
\label{eq:SU(3)} 
2g_{\omega K}=\sqrt{2}\,g_{\phi K}=2\,g_{\rho K}=\,g_{\rho \pi}=6.04 \:, 
\end{equation} 
where the value of $g_{\rho \pi}$ is due to the $\rho \to 2\pi$ decay 
width. As mentioned in Sec.~\ref{sec:intro}, this choice corresponds to  
the underlying Tomozawa-Weinberg lowest-order term in chiral perturbation 
theory \cite{Wei07}. The value of $g_{\omega K}^{\rm SU(3)}=3.02$ adopted 
here is considerably lower than the quark-model (QM) value applied to NL-SH, 
$g_{\omega K}=\frac{1}{3}g_{\omega N}^{\rm NL-SH}=4.32$, which was used in 
our previous work \cite{GFG07}, and we consider it to be the minimal value 
suggested by theory. The $\bar K$ RMF {\it vector} potential at threshold 
in nuclear matter is then given, in the static approximation, by using the 
last two terms on the rhs of Eq.~(\ref{eq:self}): 
\begin{equation} 
\label{eq:vec} 
V^{K^-}_{\rm RMF-vector}=-\frac{g_{\omega K}^{\rm SU(3)}g_{\omega N}^
{\rm NL-SH}\rho_0}{m_{\omega}^2}-{\frac{1}{2m_K}}{\left(\frac{g_{\omega K}^
{\rm SU(3)}g_{\omega N}^{\rm NL-SH}\rho_0}{m_{\omega}^2}\right)}^2=-76.7~
{\rm MeV} \,,
\end{equation} 
with $m_{\omega}=783$ MeV and $\rho_0^{\rm NL-SH}=0.146~{\rm fm}^{-3}$. 
We point out that the value $g_{\omega N}^{\rm NL-SH}=12.95$ is not far 
away from the value $g_{\omega N}^{\rm ESC04}=11.06$ from the latest 
$NN$-potential fit by the Nijmegen group \cite{Rij06}. This latter value 
was obtained in that $NN$ analysis after allowing for part of the isoscalar 
vector-meson field strength to result from a combined $\rho$-$\pi$ exchange.  
We also studied the role of isovector ${\bar K}$ nucleus interactions 
by comparing the results of using the present SU(3) choice 
$g_{\rho K}^{\rm SU(3)}=3.02$ with results applying a QM {\it universal} 
isospin coupling to NL-SH: $g_{\rho K}=g_{\rho N}^{\rm NL-SH}=4.38$. 
This value is substantially higher than the Nijmegen potential fit 
value $g_{\rho N}^{\rm ESC04}=2.77$, apparently to compensate for 
disregarding the almost four times higher value of the tensor coupling 
constant $f_{\rho N}^{\rm ESC04}$. 

SU(3) symmetry is not much of help in fixing the (anti-)kaon coupling to the
scalar-meson field $\sigma$, simply because the microscopic origin of the
$\sigma$ field and its various couplings are not unambiguous. It has been shown
recently that interpreting the $\sigma$ field in terms of a
$(J^{\pi},I)=(0^+,0)$ resonance in the $\pi\pi$-$K \bar{K}$ coupled-channel
system leads to a vanishing $\bar K N$ forward-scattering amplitude at
threshold, thus suggesting a vanishing contribution to the corresponding $\bar
K$-nucleus optical potential \cite{TKO07}. However, even for the empirically
large value of $g_{\sigma N}$ obtained in the $NN$ case ($g_{\sigma N}^{\rm
ESC04}=10.17$) and also within the RMF description of nuclei, there is no
consensus on its microscopic origin, except that QCD sum-rules do produce
strong scalar condensates. Modern $NN$ potentials using chiral perturbation
theory guidelines obtain a rather strong isoscalar-scalar two-pion exchange
contribution involving excitation of $\Delta(1232)$ in intermediate states
\cite{FKV06}. A similar two-pion exchange contribution for $\bar K N$,
involving the excitation of $K^{*}(892)$, cannot be excluded at present. In the
absence of QCD sum-rule determinations of $g_{\sigma K}$, one relies for an
order of magnitude estimate on simplified models such as the QM, giving rise to
$g_{\sigma K}^{\rm QM}=\frac{1}{3}\,g_{\sigma N}$, which for the NL-SH model gives
$g_{\sigma K}^{\mathrm QM} =3.48$. The associated RMF $K^-$ nuclear {\it
scalar} potential, in the static approximation, is given by:
\begin{equation} 
\label{eq:scalar} 
V^{K^-}_{\rm RMF-scalar}=-\frac{g_{\sigma K}^{\rm QM}g_{\sigma N}^{\rm NL-SH}\rho^s_0}
{2\, m_{\sigma}^2}=-66.3~{\rm MeV} \,, 
\end{equation} 
using the values $m_{\sigma}=526.1~{\rm MeV}$ from NL-SH and 
$\rho^s_0 \approx 0.9 \rho_0$, where $\rho^s_0$ is the scalar density. 
Our choice of $g_{\sigma K}$ is conceptually different, fitting 
$g_{\sigma K}$ to several selected values of ${\bar K}$ separation energy 
$B_{\bar K}$ in nuclear systems with a single $\bar K$ meson. These fitted values, 
all of which were considerably lower than the QM value, are specified in the 
next section. Thus, our scalar potentials are viewed as a supplement to the 
minimal vector potentials discussed above to scan over ${\bar K}$ 
nuclear binding energies in a given energy range, without imparting any 
microscopic meaning to these scalar potentials.  
We also tested the effect of adding another scalar-meson field that couples 
exclusively to strangeness, ``hidden strangeness'' $\sigma^*$ meson with mass 
$m_{\sigma^*}=980$~MeV, and coupling constant $g_{\sigma^* K} = 2.65$ 
determined from $f_0(980) \to K^+K^-$ decay \cite{SMi96}.

\section{Results and discussion}
\label{sec:res}

\subsection{Saturation of ${\bar K}$ binding energies and hadronic densities}

\begin{figure}[h!]
\includegraphics[scale=0.6]{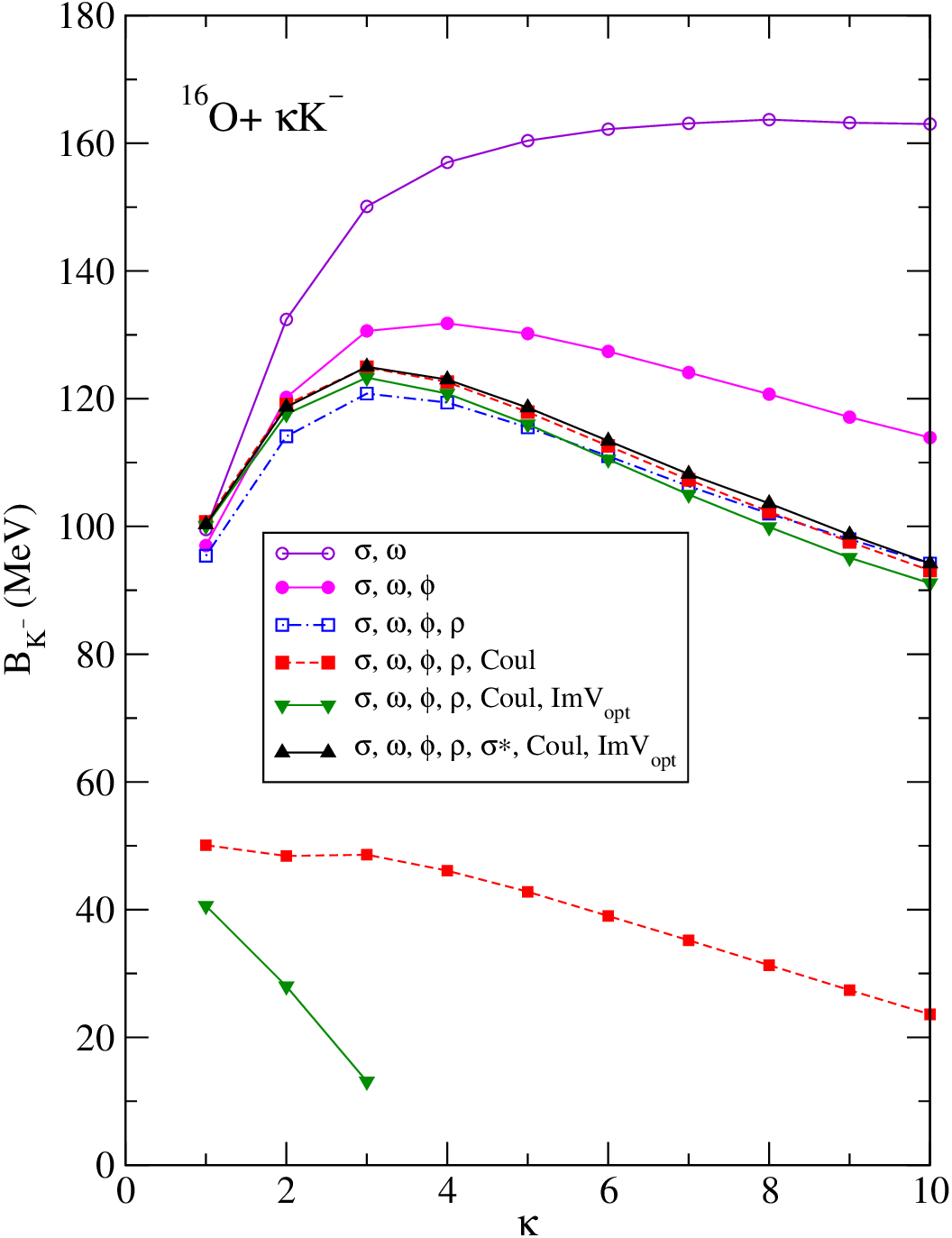}
\caption{$1s$ $K^-$ separation energy $B_{K^-}$ in $^{16}{\rm O}$+$\kappa {K^-}$
as a function of the number $\kappa$ of antikaons in several calculations 
detailed in the text, as listed in the inset, using the NL-SH RMF nuclear 
model with $g_{\sigma K}=2.433$ for the upper group of curves and 
$g_{\sigma K}=1.703$ for the lower group.}
\label{fig:f1}
\end{figure}

\begin{figure}[h!]
\includegraphics[scale=0.53]{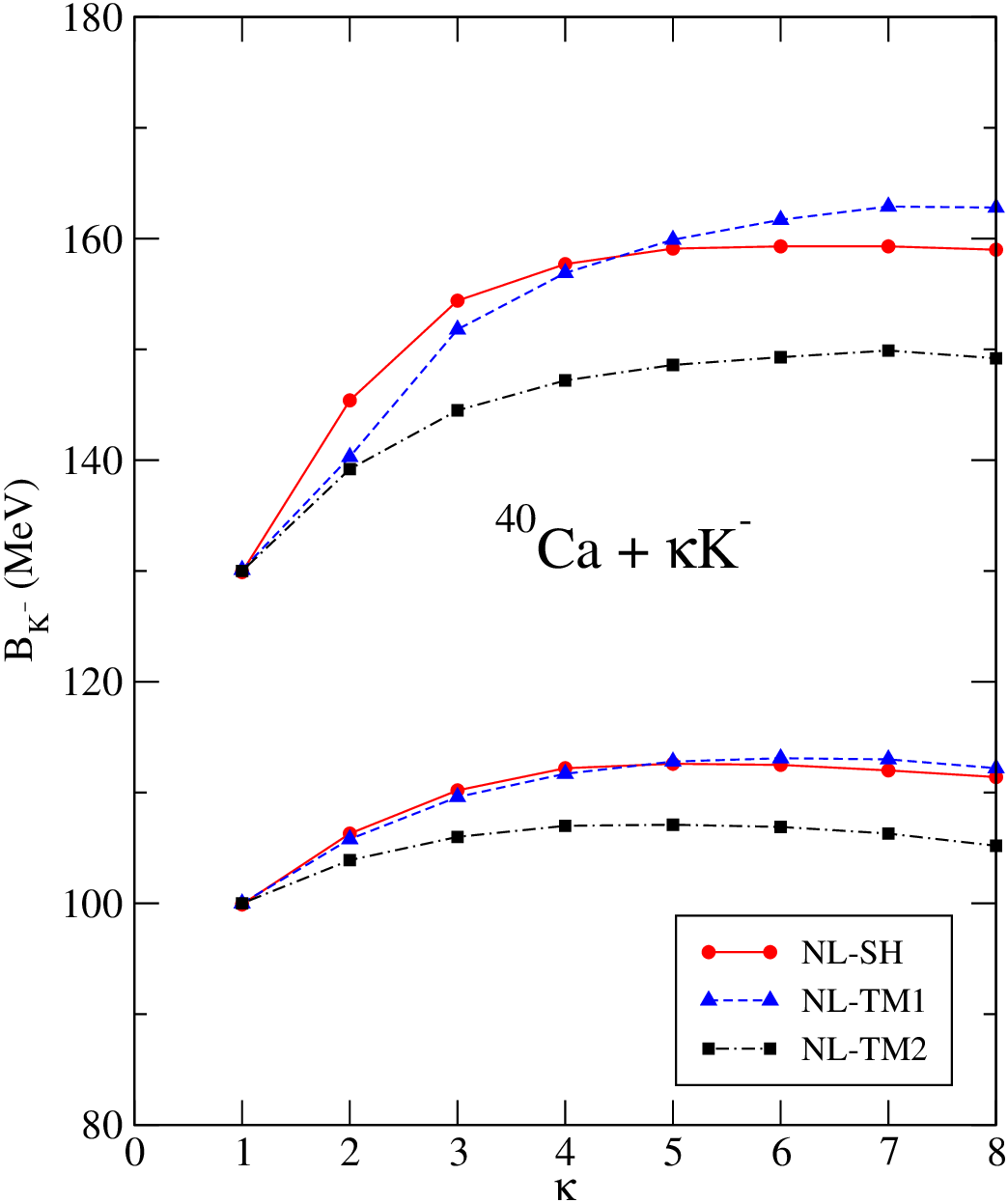}
\caption{$1s$ $K^-$ separation energy $B_{K^-}$ in $^{40}{\rm Ca}$+$\kappa {K^-}$, 
as a function of the number $\kappa$ of $K^-$ mesons, calculated in the NL-SH 
(circles, solid lines), NL-TM1 (triangles, dashed lines), and NL-TM2 
(squares, dot-dashed lines) RMF models. The lower (upper) group of curves 
was constrained to produce $B_{K^-}=100~(130)$ MeV for $\kappa=1$.} 
\label{fig:f2}
\end{figure}

Following the observation made in Ref.~\cite{GFG07} that ${\bar K}$ 
binding energies, as well as nuclear and ${\bar K}$ densities, saturate 
on increasing the number $\kappa$ of ${\bar K}$ mesons, we have explored 
how robust this saturation is. In particular we studied, for several 
selected nuclei across the periodic table, the role of various components 
of the ${\bar K}$-nucleus interaction in establishing saturation and the 
sensitivity to the choice of the RMF model. Representative examples are 
shown in Figs.~\ref{fig:f1} and \ref{fig:f2}. 

Figure~\ref{fig:f1} presents the $1s$ $K^-$ separation energy $B_{K^-}$ in 
multi-$K^-$ nuclei $^{16}{\rm O}+\kappa {K^-}$ as a function of the 
number $\kappa$ of $K^-$  mesons, using the NL-SH RMF parametrization, for 
several mean-field compositions of the $K^-$ self-energy Eq.~(\ref{eq:self}) 
with $K^-$ vector-meson couplings given by Eq.~(\ref{eq:SU(3)}). The upper 
group of curves is based on a value of $g_{\sigma K}=2.433$ ensuring 
$B_{K^-}=100$~MeV for $\kappa=1$. The $\phi,\rho,\sigma^*$ meson fields 
do not practically contribute in this case, whereas the Coulomb field adds 
a few MeV attraction and ${\rm Im}V_{\rm opt}$ adds a few MeV repulsion. 
For $\kappa > 1$, the various curves of the upper group diverge from each 
other: with respect to a ``minimal'' $\sigma$+$\omega$ model (open circles), 
the main contributors are the repulsive $\phi$ and $\rho$ {\it vector} mesons, 
as judged by the curves marked by solid circles and open squares, 
respectively. Given their contributions, which get larger with $\kappa$, 
the inclusion of the Coulomb field, the $\sigma^*$ meson field and 
${\rm Im}V_{\rm opt}$ makes a small difference. 
However, the $K^-$ absorptivity ${\rm Im}V_{\rm opt}$ makes a big 
difference for the lower group of curves consisting of only two choices, 
both with $g_{\sigma K}=1.703$ fitted to $B_{K^-}\hspace{-2pt} \approx 40-50$~MeV for 
$\kappa=1$. The energy dependence of ${\rm Im}V_{\rm opt}$ magnifies its 
effect for relatively low values of $B_{K^-}$ in the region 
$B_{\bar K}\leq 100$~MeV, adding significant repulsion the lesser the value 
of $B_{K^-}$ is. This added repulsion (lowest curve in Fig.~\ref{fig:f1}) 
leads to a rapid fall-off of $B_{K^-}$, terminating the binding at $\kappa=3$, 
because the system $^{16}{\rm O}$+$4 {K^-}$ is found to be unbound for this 
particular choice of $\kappa = 1$ parameters. 
The lesson from Fig.~\ref{fig:f1} is that saturation of the $K^-$ binding 
energy in nuclear systems with $\kappa$ $K^-$mesons is a robust phenomenon, 
which remains valid regardless of the type of meson fields mediating the 
strong interaction among antikaons and nucleons, provided a {\it minimal} 
isoscalar vector-meson field ($\omega$) is included. For a sufficiently 
large number $\kappa$ of $K^-$ mesons, the combined repulsive $K^-K^-$ 
interaction generated by the vector meson fields $\omega,\:\phi,\:\rho$ 
wins over the attractive interaction generated by the isoscalar scalar-meson fields (dominated by $\sigma$). The effect of adding the $\sigma^*$ 
scalar field is found to be insignificant. These conclusions hold also 
for a Lagrangian in which scalar fields are introduced differently than 
in Eq.~(\ref{eq:lagrangian}), resulting in a correspondingly different 
definition of effective masses, 
$m_K^*=m_K-g_{\sigma K}\sigma_0-g_{\sigma^* K}\sigma^*_0$ \cite{GSB98}, 
than in Eq.~(\ref{eq:meff}). 

Figure~\ref{fig:f2} shows the $1s$ $K^-$ separation energy $B_{K^-}$ in 
multi-${K^-}$ nuclei $^{40}{\rm Ca}+\kappa {K^-}$ as a function of the 
number $\kappa$ of $K^-$ mesons, calculated in the NL-SH, NL-TM1, and 
NL-TM2 RMF models for two choices of $g_{\sigma K}$ designed, within each 
model, to produce $B_{K^-}=100$ and 130 MeV for $\kappa=1$. The values of 
$g_{\sigma K}$ for NL-SH were 1.703 and 2.993, respectively. The difference 
between the various curves, for a given starting value of $B_{K^-}$, 
originates from the specific balance in each one of these RMF models between 
the vector fields and the scalar field. The figure illustrates that the 
saturation of the ${\bar K}$ binding energy in nuclear systems with several 
antikaons is not limited to a particular choice of RMF parametrization
but is a general feature independent of the applied RMF model. Without loss 
of generality, we therefore specialize in the subsequent discussion to a
specific RMF model, namely NL-SH. 

\begin{figure}[h!]
\includegraphics[scale=0.6]{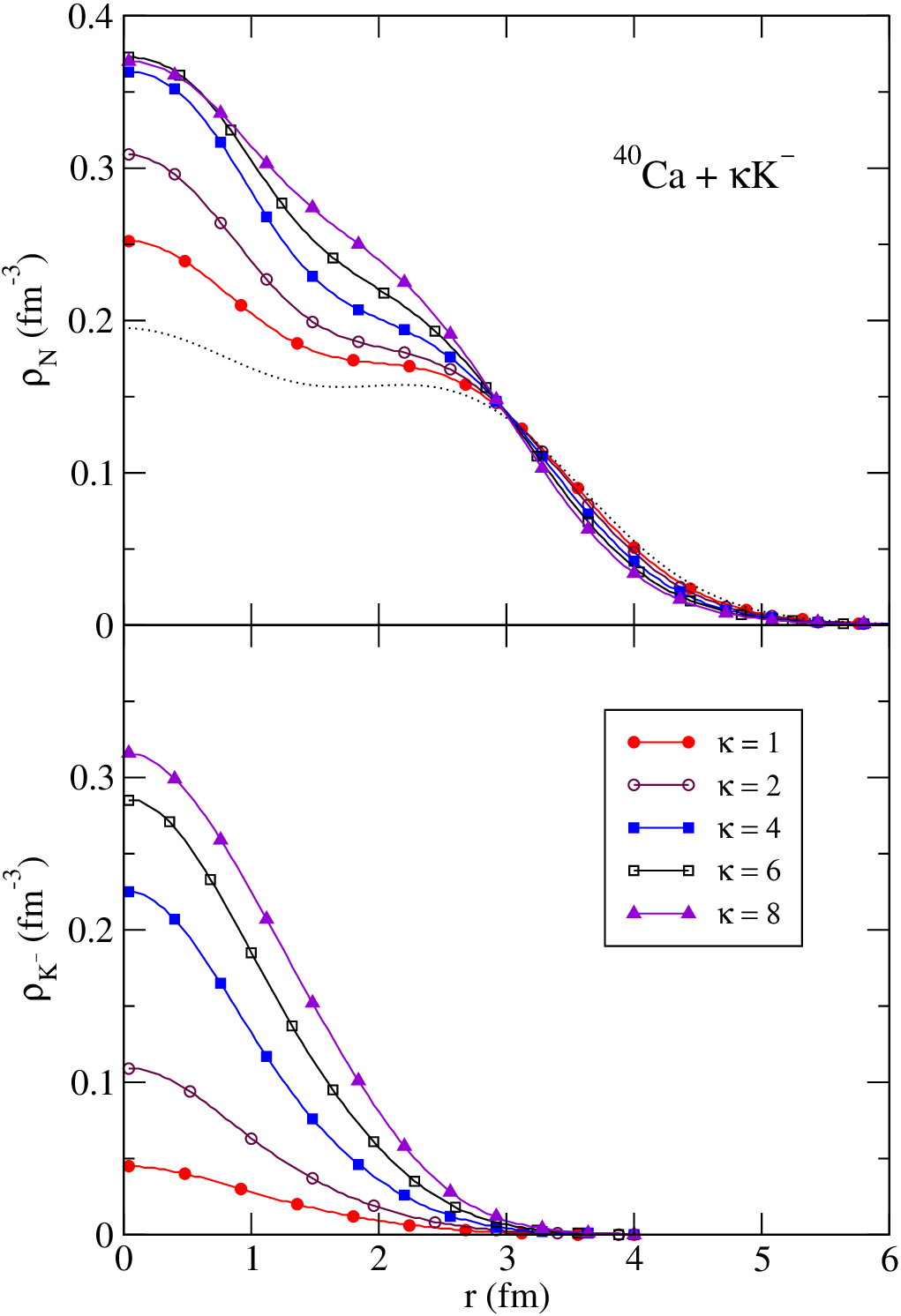}
\caption{Nuclear density $\rho_N$ (top panel) and $1s$ ${K^-}$ density
$\rho_{K^-}$ (bottom panel) in $^{40}{\rm Ca}$+$\kappa K^-$, calculated in 
the NL-SH RMF model, with $g_{\sigma K}=1.703$ chosen to yield 
$B_{K^-}=100$~MeV in $^{40}{\rm Ca}$+$1 K^-$. The dotted curve stands for 
the $^{40}$Ca density in the absence of $K^-$ mesons.}
\label{fig:f3}
\end{figure}

\begin{figure}[h!]
\includegraphics[scale=0.6]{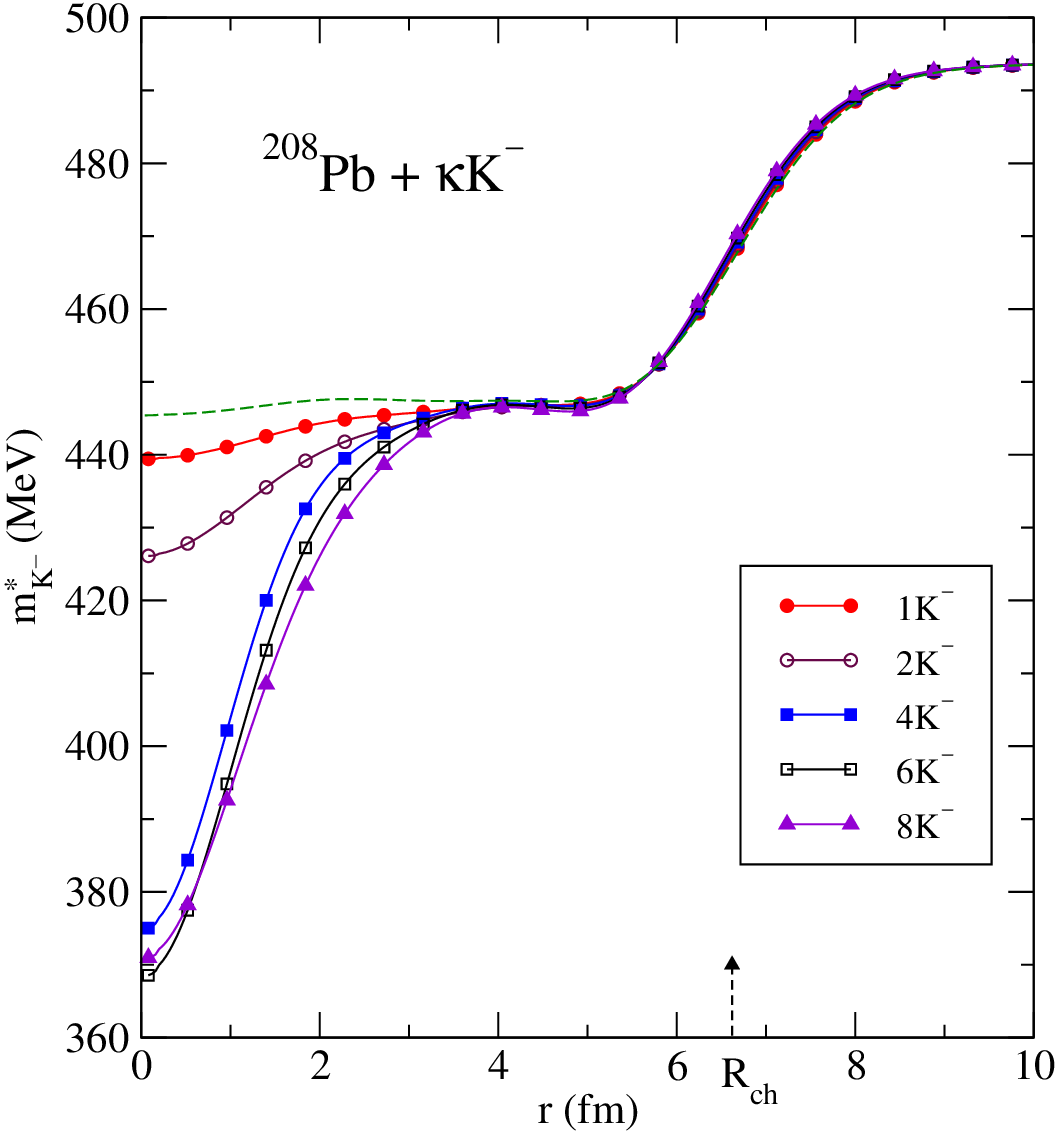}
\caption{$1s$ $K^-$ effective mass $m^*_{K^-}$ in multi-${K^-}$ nuclei 
$^{208}{\rm Pb}$+$\kappa K^-$, calculated in the NL-SH  model with 
$g_{\sigma K}=2.433$ chosen to yield $B_{K^-}=100$~MeV in 
$^{208}{\rm Pb}$+$1K^-$. The dashed curve stands for the ``static'' case where 
the $^{208}{\rm Pb}$ $\sigma$ field in a purely nuclear RMF calculation was 
used in Eq.~(\protect{\ref{eq:meff}}) for $m^*_{K^-}$. The dashed arrow 
indicates the charge half-density radius $R_{\rm ch}$ in $^{208}{\rm Pb}$.} 
\label{fig:f4}
\end{figure}

The dependence of the nuclear density $\rho_N(r)$ and the $K^-$
density $\rho_{K^-}(r)$ on the number $\kappa$ of $K^-$ mesons
in multi-${K^-}$ nuclei $^{40}{\rm Ca}+\kappa {K^-}$ is shown in 
Fig.~\ref{fig:f3}. The coupling constant $g_{\sigma K}=1.703$ was chosen 
such that the single-$K^-$ configuration was bound by 100~MeV, the same 
as for the NL-SH lower curve in Fig.~\ref{fig:f2}. The density distribution 
$\rho_N$ for $^{40}$Ca is also shown, for comparison, by the dotted curve 
in the upper panel. It is clear from this figure that the central nuclear 
density $\rho_N$ saturates for $\kappa=8$ at a value about twice larger than 
that for $\rho_N$ in $^{40}$Ca. In the lower panel, it is seen that the 
gradual increase of $\rho_{K^-}$ with $\kappa$ slows down with increasing 
$\kappa$. 

The saturation of the nuclear density in multi-$\bar K$ nuclei manifests 
itself also in the behavior of the ${\bar K}$ effective mass in the nuclear medium, 
Eq.~(\ref{eq:meff}), as a function of the number $\kappa$ 
of antikaons. This is illustrated for $^{208}{\rm Pb} + \kappa K^-$ in 
Fig.~\ref{fig:f4}. Note that the calculated effective mass distribution 
$m^*_{K^-}(r)$ remains almost independent of the number of $K^-$ mesons 
over a large volume of the nucleus, for $r \geq 3$--4~fm, reflecting 
a similar $\kappa$ independence of the scalar $\sigma$ field through the 
underlying nuclear density. In fact, the $\sigma$ field in this region 
is almost unaffected by the presence of $K^-$ mesons, as demonstrated 
by the dashed curve which uses the ``static'' $^{208}{\rm Pb}$ $\sigma$ 
field from a purely nuclear RMF calculation.  
It is only within a relatively small region near the nuclear center, 
typically $r \leq 2$--3~fm, that the variation of $m^*_{K^-}(r)$ with
$\kappa$ gets to be more pronounced. However, $m^*_{K^-}(r=0)$ quickly 
saturates, already for $\kappa \approx 8$. The figure demonstrates clearly 
that the concept of ``nuclear matter'' is far from being realized, even for 
a nucleus as large as $^{208}{\rm Pb}$ and that conclusions made on 
$\bar K$ binding and kaon condensation in finite nuclei, using nuclear-matter 
arguments, should be taken with a grain of salt.

\subsection{Exotic ${\bar K}$ nuclear configurations} 

\begin{figure}[h!]
\includegraphics[scale=0.6]{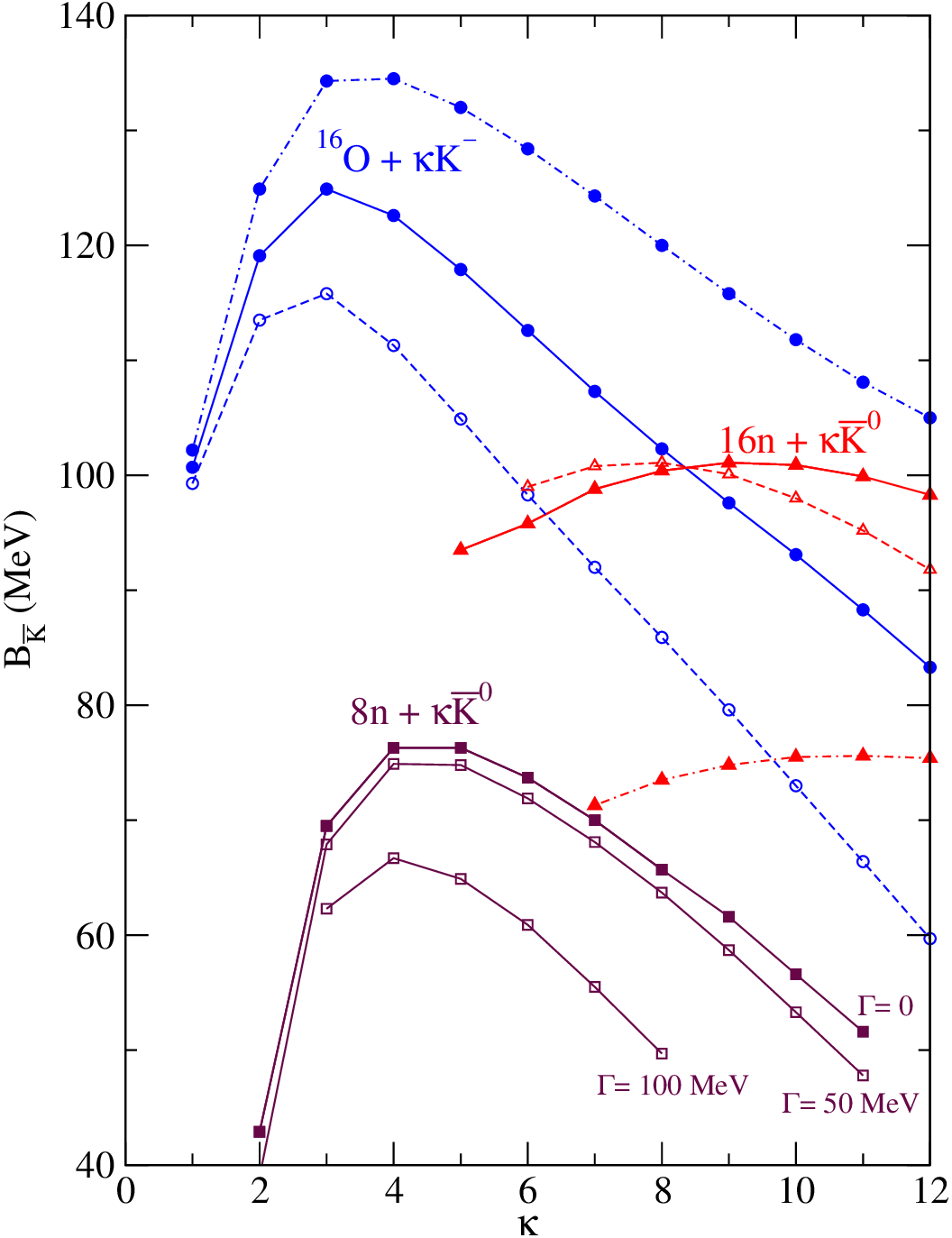}
\caption{$1s$ $\bar K$ separation energy $B_{\bar K}$ in $^{16}{\rm O}$+$
\kappa K^-$, $16n$+$\kappa {\bar K}^0$, and $8n$+$\kappa {\bar K}^0$, 
as a function of $\kappa$, calculated in the NL-SH RMF model, with 
$g_{\rho K}=0$ (dot-dashed curves), $g_{\rho K}^{\rm SU(3)} = 3.02$ 
(solid curves) and $g_{\rho K} = g_{\rho N} = 4.38$ (dashed curves). 
ImV$_{\rm opt}=0$ is assumed everywhere except for the two lowest solid curves 
(open squares) in $8n$+$\kappa {\bar K}^0$ where ImV$_{\rm opt} \neq 0$, 
such that the value of width $\Gamma_{{\bar K}^0}$ is held fixed at 50 
and 100 MeV, respectively, see text.} 
\label{fig:f5}
\end{figure}

Because in the underlying $\bar K N$ dynamics the $I=0$ interaction is 
considerably more attractive than the $I=1$ interaction, we have looked for 
ways to maximize the role of the $\bar K N~I=0$ channel in multi-$\bar K$ 
nuclei. For a nuclear core with $N=Z$, no matter which charge states are 
assigned to the $\bar K$ mesons, the average $\bar K N$ interaction is given 
by a $(2I+1)$-average which disfavors the $I=0$ channel. This disadvantage 
is partly removed by considering $n{\bar K}^0$ (or $pK^-$) multi-$\bar K$ 
nuclei, where both isospin channels assume equal weight, so that the 
stronger $I=0$ component may provide sufficient attraction to overcome 
the insufficient attraction in purely neutron matter. We therefore studied 
exotic configurations consisting solely of ${\bar K}^0$ mesons bound 
to neutrons. Our calculations confirmed that ${\bar K}$ mesons 
can bind together systems of nucleons that otherwise are unbound. 

In Fig.~\ref{fig:f5}, we compare the separation energies $B_{\bar K}$ in 
$16n$+$\kappa {\bar K}^0$ and in $8n$+$\kappa {\bar K}^0$ exotic multi-$\bar K$ 
configurations with $B_{\bar K}$ in $^{16}{\rm O}$+$\kappa K^-$ multi-$\bar K$ nuclei, 
most of which were calculated in the NL-SH RMF model with the ``canonical'' 
$g_{vK}$ coupling constants of Eq.~(\ref{eq:SU(3)}) and $g_{\sigma K}=2.433$ 
chosen to yield $B_{K^-} = 100$~MeV in $^{16}{\rm O}$+$1 K^-$ as in 
Fig.~\ref{fig:f1}, and for ImV$_{\rm opt}=0$. For each sequence of 
multi-$\bar K$ nuclei, $B_{\bar K}$ increases as a function of $\kappa$ 
to a maximum value and then starts to decrease. Whereas the sequence 
consisting of 8 neutrons plus ${\bar K}^0$ mesons starts with $\kappa=1$ 
(not shown in the figure), a larger number of neutrons generally requires 
a threshold value for $\kappa$ as shown for the sequences consisting of 16 
neutrons plus ${\bar K}^0$ mesons. Exceptions to the use of the canonical 
$g_{vK}$ set of Eq.~(\ref{eq:SU(3)}), or to ImV$_{\rm opt}=0$, are as 
follows: 
\begin{itemize} 

\item $g_{\rho K}^{\rm SU(3)} = 3.02$ was used everywhere except for 
$g_{\rho K}=0$ in the dot-dashed curves and except for
$g_{\rho K} = g_{\rho N} = 4.38$ 
(universal isospin coupling) in the dashed curves to study the role 
of the $\rho$ meson in ``nonexotic'' multi-$\bar K$ nuclei 
($^{16}{\rm O}+\kappa K^-$) and in ``exotic'' ones ($16n$+$\kappa {\bar K}^0$). 
In $^{16}{\rm O}$+$\kappa K^-$, the values of $B_{K^-}$ for a given value of 
$\kappa > 1$ decrease as $g_{\rho K}$ is increased, as expected from the 
{\it repulsive} $K^-K^-$ isovector interaction. In contrast, the larger 
the value of $g_{\rho K}$ is, the larger is the value of $B_{{\bar K}^0}$ 
expected in $16n$+$\kappa {\bar K}^0$, because it is the $\rho$ isovector 
interaction that distinguishes the more attractive $I=0$ component of the 
${\bar K}^0 n$ interaction from the less attractive $I=1$ component. 
Indeed, this holds for $\kappa \leq 8$ in the figure. 
However, for $\kappa > 8$, the contribution of the repulsive $\bar{K}^0\bar{K}^0$ 
isovector interaction becomes substantial for the values of 
$g_{\rho K} \neq 0$ used here; the $B_{{\bar K}^0}$ dashed curve for 
the universal $\rho$ coupling heads down, crossing the  $B_{{\bar K}^0}$ 
solid curve corresponding to SU(3) $\rho$ coupling. All in all, substantial 
binding in $16n$+$\kappa {\bar K}^0$ multi-$\bar K$ nuclei is reached 
for these values of $g_{\rho K} \neq 0$. 

\item The effect of ImV$_{\rm opt}$ on $B_{\bar K}$ is relatively 
unimportant for $B_{\bar K}\geq 100$~MeV, where the dominant $\bar K N \to 
\pi \Sigma$ decay channel is closed. The inclusion of ImV$_{\rm opt}$ is 
found then to induce repulsion of less than 5 MeV. 
However, in $8n$+$\kappa {\bar K}^0$ multi-$\bar K$ nuclei, 
where $B_{{\bar K}^0}\leq 80$~MeV, the effect of ImV$_{\rm opt}$ becomes 
significant. An estimate of this effect is given by comparing the 
$B_{{\bar K}^0}$ curve for $\Gamma=0$ (solid squares) with the 
$B_{{\bar K}^0}$ curves using ImV$_{\rm opt} \neq 0$ (open squares) 
such that the value of $\Gamma_{{\bar K}^0}$ is {\it held fixed} at 50 
and 100 MeV. As expected, the larger input widths induce a stronger 
repulsion that lowers the calculated $B_{{\bar K}^0}$ values. Yet
considerably lower values of $B_{{\bar K}^0}$ are obtained once the 
${\bar K}^0$ widths are included self-consistently in these dynamical 
calculations. 
\end{itemize}

\begin{figure}[h!]
\includegraphics[scale=0.6]{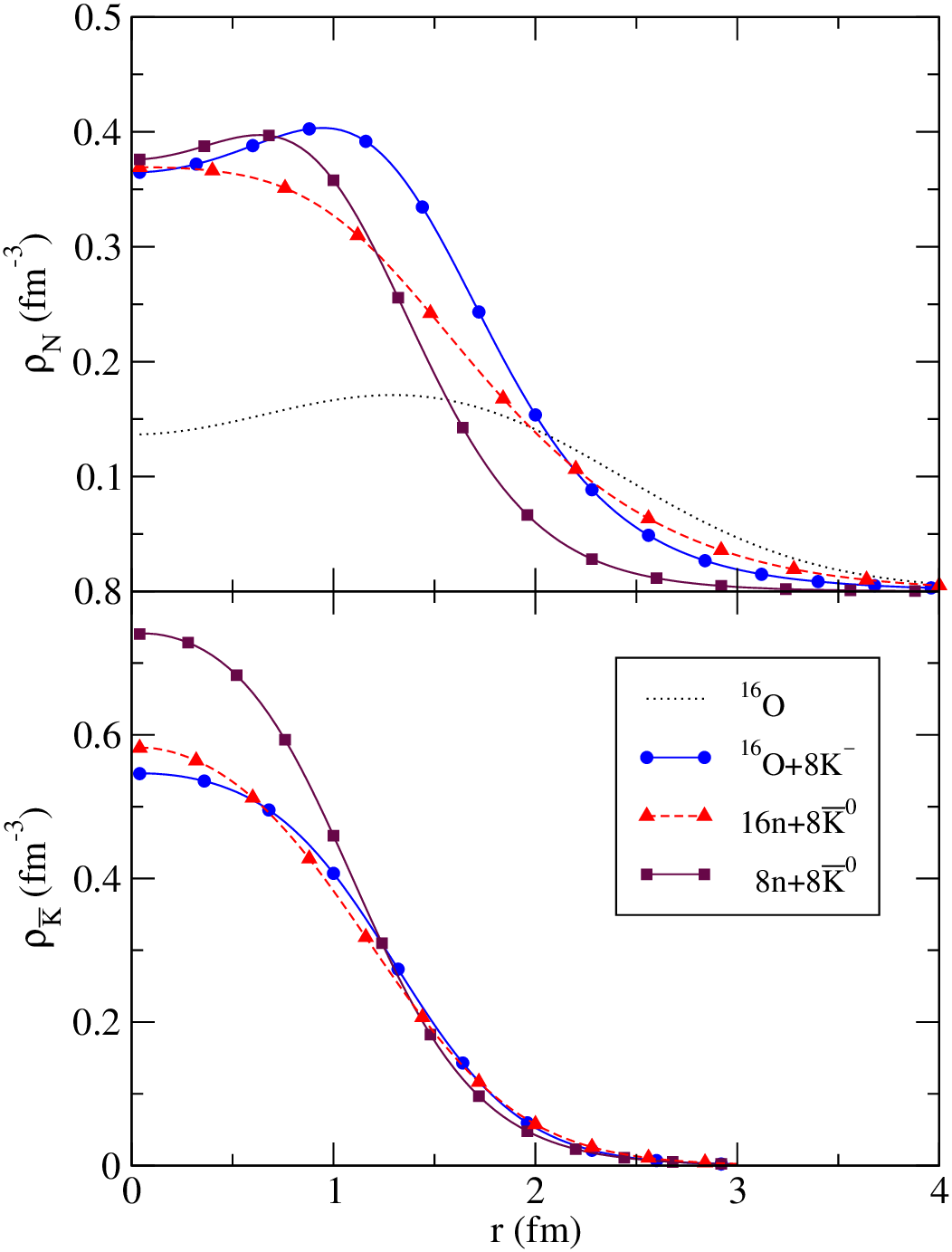}
\caption{Nuclear density $\rho_N$ (top panel) and $1s$ ${\bar K}$ density
$\rho_{\bar K}$ (bottom panel) in $^{16}{\rm O}$+$8K^-$, $16n$+$8{\bar K}^0$ 
and $8n$+$8{\bar K}^0$, calculated in the NL-SH RMF model, with the 
``canonical'' $g_{vK}$ coupling constants of Eq.~({\protect{\ref{eq:SU(3)}}}) 
and with $g_{\sigma K} = 2.433$ to yield $B_{K^-} = 100$~MeV in 
$^{16}{\rm O}$+$1 K^-$ as in Fig.~\ref{fig:f5}. The dotted curve stands 
for the $^{16}$O density in the absence of $\bar K$ mesons.}
\label{fig:f6}
\end{figure}

The nucleon-density distribution $\rho_N(r)$ and the $\bar K$-density
distribution $\rho_{\bar K}(r)$ are shown in Fig.~\ref{fig:f6} for $^{16}{\rm
O}$+$8 K^-$, $16n$+$8 {\bar K}^0$, and $8n$+$8 {\bar K}^0$. We note that
$\rho_N$ and $\rho_{\bar K}$ are normalized to the number of nucleons and
number of antikaons, respectively. The $\bar K$ couplings were chosen such that
the 1$K^-$ configuration in $^{16}{\rm O}$ is bound by 100~MeV, as in
Fig.~\ref{fig:f5}. For comparison, we also present the density distribution
$\rho_N$ for $^{16}$O without $\bar K$ mesons. Owing to the substantial $\bar
K$ density $\rho_{\bar K}$ in the nuclear center, the central nuclear density
$\rho_N(0)$ in all three systems with 8 $\bar K$ mesons is about 2-3 times
larger than the central nuclear density $\rho_0$ in $^{16}$O for $\kappa=0$.
The situation is particularly pronounced in $8n$+$8{\bar K}^0$, with the same
central density $\rho_N(0)$ as in the systems with 16 nucleons + 8$\bar K$.
Furthermore, the $8n$+$8{\bar K}^0$ system is compressed substantially in
comparison with the other multi-$\bar K$ systems, judging by the radial
extension of $\rho_N$ and $\rho_{\bar K}$ in both panels of Fig.~\ref{fig:f6}.
The relatively high value  $\rho_{\bar K}(0) \sim 5 \rho_0$ for this system
does not introduce complications due to possible overlap between antikaons,
because the mean-square radius of $K^-$ is less than half of the corresponding
quantity for the proton \cite{Ame86,SSB80}.

\begin{figure}[h!]
\includegraphics[scale=0.6]{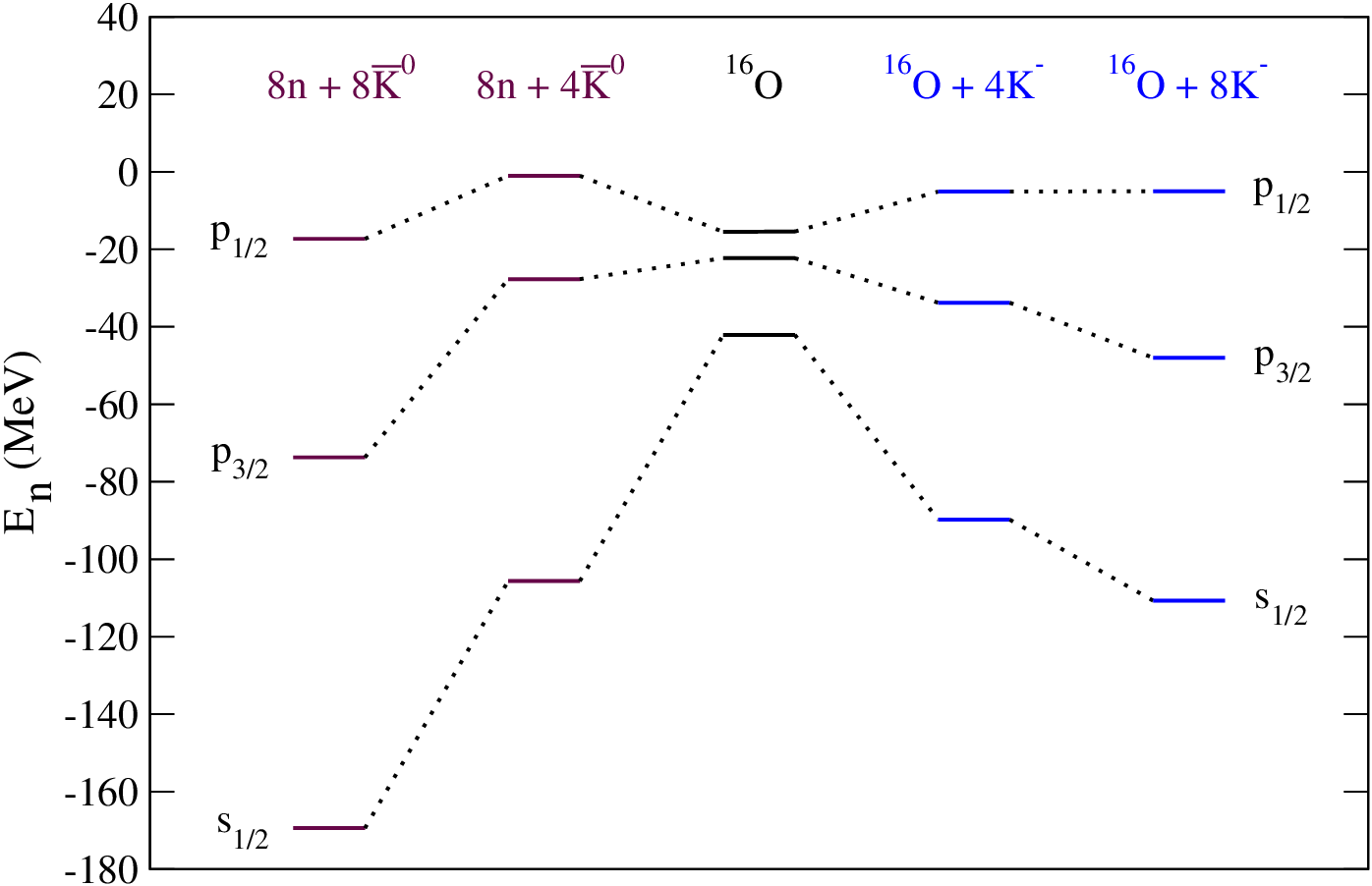}
\caption{Neutron single-particle spectra in $^{16}{\rm O}$ (center) 
$^{16}{\rm O}$+$4(8)K^-$ (right) and $8n$+$4(8){\bar K}^0$ (left), calculated 
in the NL-SH RMF model with $g_{\sigma K}=2.433$ chosen to yield 
$B_{K^-}=100$~MeV in $^{16}{\rm O}$+$1 K^-$.}
\label{fig:f7} 
\end{figure} 

Given the compressed nuclear densities plotted in Fig.~\ref{fig:f6}, we show in
Fig.~\ref{fig:f7} the calculated neutron single-particle energy levels in
$^{16}{\rm O}$, in $^{16}{\rm O}$+$\kappa K^-$ and in $8n$+$\kappa {\bar K}^0$
multi-$\bar{K}$ nuclei for $\kappa=4,~8$. The $\bar K$ couplings are the same
as in Figs.~\ref{fig:f5} and \ref{fig:f6}, again chosen to ensure
$B_{K^-}=100$~MeV in $^{16}{\rm O}$+$1 K^-$. The $1s_{1/2}$ and $1p_{3/2}$
levels undergo increasingly attractive shifts on varying $\kappa$ in these
multi-$\bar{K}$ systems. Particularly strong is the downward shift of the
$1s_{1/2}$ level, by about 70~MeV in $^{16}{\rm O}$+$8K^-$ and by about 130~MeV
in $8n$+$8{\bar K}^0$. In contrast, the $1p_{1/2}$ neutron level is pushed up by
about 10~MeV in the $^{16}{\rm O}$+$\kappa K^-$ systems as a result of a
gradually increasing spin-orbit splitting which reaches 43~MeV for $\kappa = 8$
(recall that it is 7 MeV for $\kappa = 0$ using NL-SH). The $1p_{1/2}$ neutron
level is weakly bound in the exotic $8n$+${\kappa}{\bar K}^0$ systems for $1 < \kappa
< 6$, getting more bound with $\kappa$ as shown in the figure for these
systems. The $1p$ spin-orbit splitting becomes as large as 56~MeV in the exotic
$8n+8{\bar K}^0$ system which exhibits the largest single-particle level
splittings in this figure. Here the $1p_{1/2}$ neutron level, too, undergoes
attraction. 

\begin{figure}[h!]
\includegraphics[scale=0.6]{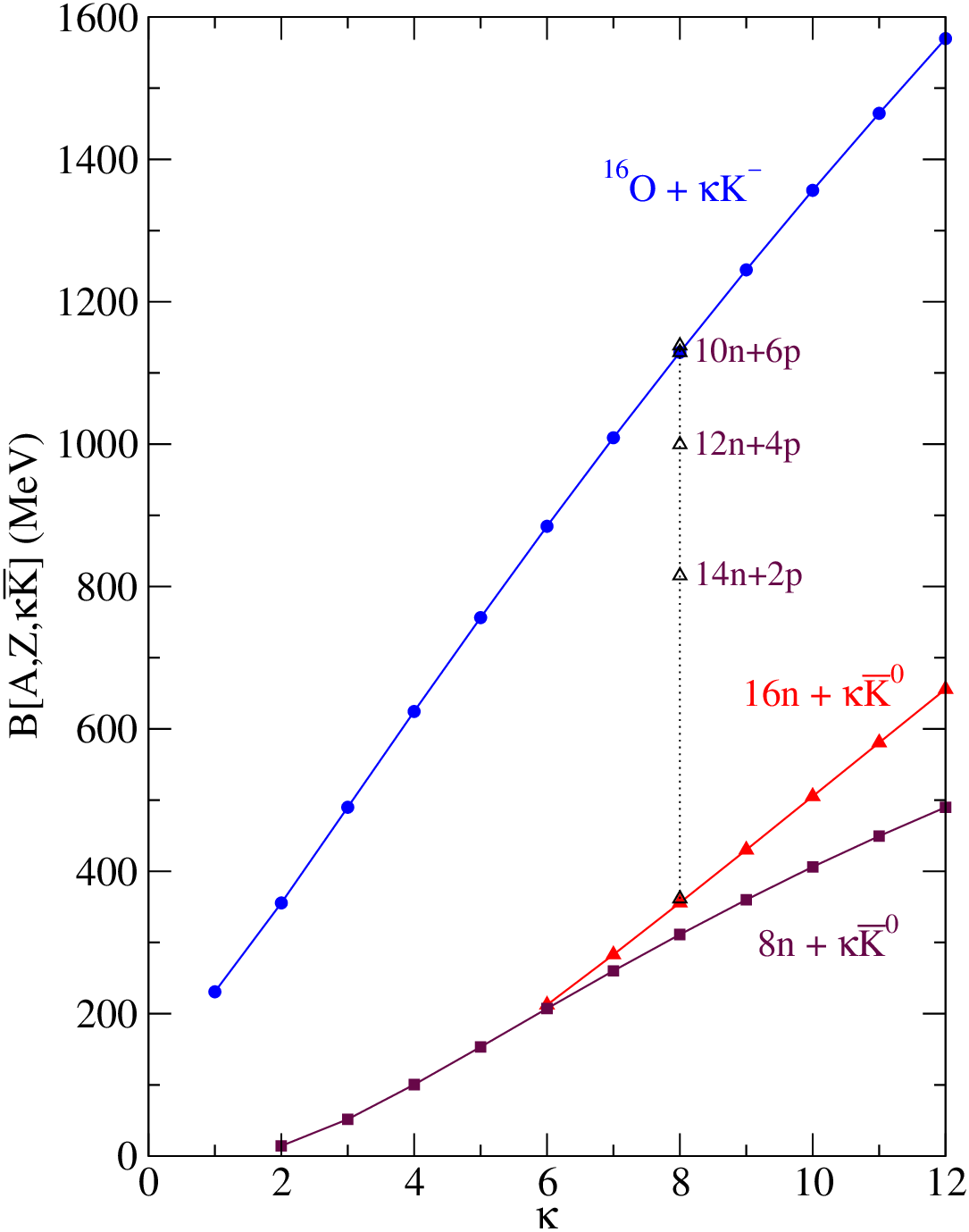}
\caption{Total binding energy $B[A,Z,\kappa{\bar K}]$ of 
$^{16}{\rm O}$+$\kappa K^-$ (circles), $16n$+$\kappa {\bar K}^0$ (solid triangles) 
and  $8n+\kappa {\bar K}^0$ (squares) multi-$\bar{K}$ systems, as a function 
of $\kappa$, calculated in the NL-SH RMF model with $g_{\sigma K}=2.433$ 
chosen to yield $B_{K^-}=100$~MeV in $^{16}{\rm O}$+$1 K^-$, and with 
$g_{\rho K}=0$. For $\kappa=8$, total binding energy values for configurations 
that are intermediate between $^{16}{\rm O}$+$8K^-$ and $16n$+$8{\bar K}^0$ are 
shown in open triangles along the dashed line.} 
\label{fig:f8} 
\end{figure}

It is worth mentioning that exotic multi-${\bar K}^0$ configurations that contain 
no protons lie high in the continuum of ``nonexotic'' multi-$\bar{K}$ nuclei 
that are based on nuclear cores with protons and neutrons. Figure~\ref{fig:f8} 
shows the calculated {\it total} binding energy $B[A,Z,\kappa {\bar K}]$, 
Eq.~(\ref{eq:sep}), assuming for simplicity $g_{\rho K}=0$, for three 
sequences of multi-$\bar{K}$ nuclei. To illustrate the relationship 
between ``exotic'' and ``nonexotic'' configurations, we take the 
$16n$+$8 {\bar K}^0$ configuration, specifically in its lowest isospin 
$I=4$ state, and replace successively ${\bar K}^0$+$n$ pairs by $K^-$+$p$ 
pairs until $^{16}{\rm O}$+$8K^-$ is reached. This is demonstrated by the 
empty triangles along the vertical dotted line that connects the initial 
and final configurations. Both initial and final configurations have 
identical quantum numbers $B=8,~Q=0,~I=4$, so they are commensurate. 
Therefore, although $\bar K$ mesons are capable of stabilizing purely 
neutron configurations, these exotic configurations do not compete 
energetically with multi-$\bar{K}$ nuclei based on nuclear cores along 
the nuclear valley of stability.

\section{Summary} 
\label{sec:concl}

In the main part of this work, we studied several dynamical aspects of 
multi-$\bar K$ nuclear states within RMF methodology. In particular, 
we discussed in detail the saturation pattern of $\bar{K}$ separation 
energies and nuclear densities on increasing the number of antikaons 
embedded in the nuclear medium. Saturation was demonstrated to be a robust 
feature of multi-$\bar K$ nuclei. The saturated values of $B_{\bar K}$, for 
``natural'' values of meson-field coupling constants were found generally to 
be below 200 MeV, considerably short of the threshold value $\approx\hspace{-4pt} 320$ 
MeV needed for the onset of kaon condensation under laboratory conditions. 
We conclude, consistently with our earlier conjecture \cite{GFG07}, that 
$\bar{K}$ mesons do not provide the physical ``strangeness'' degrees of 
freedom for self-bound strange dense matter. 

We first explored contributions of specific meson mean fields to the
$\bar{K}$ separation energy $B_{\bar K}$. Saturation of $B_{\bar{K}}$
emerged for any boson-field composition that includes the dominant 
vector $\omega$-meson field, using the ``minimal'' SU(3) value suggested 
by the leading-order Tomozawa-Weinberg term of the meson-baryon 
effective Lagrangian. Moreover, the contribution of each one 
of the vector $\phi$-meson and $\rho$-meson fields was found to be 
substantially repulsive for systems with a large number of antikaons, 
reducing the $\bar{K}$ separation energy as well as lowering the threshold 
value of number of antikaons required for saturation to occur. In contrast, 
the Coulomb interaction and the addition of a hidden-strangeness scalar 
$\sigma^*$-meson field have little effect on the binding energy balance 
and on the pattern of saturation.

We also verified that the saturation behavior of $B_{\bar K}$ is 
qualitatively independent of the RMF model applied to the nucleonic 
sector. The onset of saturation was found to depend on the atomic number. 
Generally, the heavier the nucleus is, the more antikaons it takes to 
saturate their separation energies.

The saturation phenomenon found for the $\bar{K}$ separation energy 
is also reflected in the nucleon and antikaon density distributions, 
with the assertions made above remaining valid. The saturation of the 
nuclear density in multi-$\bar{K}$ nuclei manifests itself in the behavior 
of the $\bar{K}$ effective mass distribution in the nuclear medium. 
We stress that in the case of antikaons the concept of nuclear matter 
is far from being realized even in such a heavy nucleus as $^{208}$Pb. 
Specifically, the reduction of $m^*_{\bar K}(r)$ on adding $\bar{K}$ 
mesons is pronounced only within a small region around the nuclear center. 

In the second part of this work, we studied exotic configurations 
consisting exclusively of neutrons and ${\bar K}^0$ mesons. 
We demonstrated that a finite number of neutrons can be made self-bound 
by adding few ${\bar K}^0$ mesons, with the resulting nuclear configurations 
more tightly bound than ordinary nuclei. Saturation of $B_{{\bar K}^0}$ 
was found for these exotic configurations too. Yet, these exotic 
configurations consisting exclusively of neutrons and ${\bar K}^0$ mesons 
lie high in the continuum of the less exotic multi-$\bar{K}$ configurations 
based on nuclear cores along the nuclear valley of stability. 

In conclusion, over a wide range of variations our calculations do not 
indicate any precursor phenomena to kaon condensation in self-bound 
strange nuclear systems.

\begin{acknowledgments}
This work was supported in part by the GA AVCR grant IAA100480617
and by the Israel Science Foundation grant 757/05. 
\end{acknowledgments}


\begin{thebibliography}{99} 

\bibitem{KNe86} D.B. Kaplan, A.E. Nelson, Phys. Lett. B {\bf 175}, 57 (1986); 
{\it ibid} {\bf B179}, 409 (1986). 

\bibitem{NKa87} A.E. Nelson, D.B. Kaplan, Phys. Lett. B {\bf 192}, 193 (1987).

\bibitem{FGa07} E.~Friedman, A.~Gal, Phys. Rep. {\bf 452}, 89 (2007); 
and earlier references cited therein. 
 
\bibitem{SBD06} W.~Scheinast {\it et al.}, Phys. Rev. Lett. {\bf 96}, 072301 
(2006); and earlier references to nucleus-nucleus experiments cited therein. 

\bibitem{DMo77} C.B.~Dover, P.J.~Moffa, Phys. Rev. C {\bf 16}, 1087 (1977). 

\bibitem{Lee96} C.-H.~Lee, Phys. Rep. {\bf 275}, 255 (1996). 

\bibitem{PBP97} M.~Prakash, I.~Bombaci, M.~Prakash, P.J.~Ellis, J.M.~Lattimer, 
R.~Knorren, Phys. Rep. {\bf 280}, 1 (1997). 

\bibitem{HHJ00} H.~Heiselberg, M.~Hjorth-Jensen, Phys. Rep. {\bf 328}, 237 
(2000). 

\bibitem{HPa00} H.~Heiselberg, V.R.~Pandharipande, Annu. Rev. Nucl. Part. Sci. 
{\bf 50}, 481 (2000). 

\bibitem{Gle01} N.K.~Glendenning, Phys. Rep. {\bf 342}, 393 (2001). 

\bibitem{RSW01} A.~Ramos, J.~Schaffner-Bielich, J.~Wambach, Lect. Notes 
Phys. {\bf 578}, 175 (2001). 

\bibitem{SMi96} J.~Schaffner, I.N.~Mishustin, Phys. Rev. C {\bf 53}, 1416 
(1996).

\bibitem{EKP95} P.J.~Ellis, R.~Knorren, M.~Prakash, Phys. Lett. B {\bf 349}, 
11 (1995). 

\bibitem{YDA04} T.~Yamazaki, A.~Dot\'{e}, Y. Akaishi, Phys. Lett. B {\bf 587}, 
167 (2004). 

\bibitem{SBG00} J.~Schaffner-Bielich, A.~Gal, Phys. Rev. C {\bf 62}, 034311 
(2000); and references cited therein. 

\bibitem{GFG07} D.~Gazda, E.~Friedman, A.~Gal, J.~Mare\v{s}, Phys. Rev. C 
{\bf 76}, 055204 (2007). 

\bibitem{MFG05} J. Mare\v{s}, E. Friedman, A. Gal, Phys. Lett. B {\bf 606}, 
295 (2005). 

\bibitem{MFG06} J. Mare\v{s}, E. Friedman, A. Gal, Nucl. Phys. A {\bf 770}, 
84 (2006). 

\bibitem{Wei07} W. Weise, {\it Proc. IX Int. Conf. Hypernuclear and Strange 
Particle Physics}, Eds. J.~Pochodzalla and Th.~Walcher (SIF and 
Springer-Verlag, Berlin Heidelberg, 2007) p. 243 [arXiv:nucl-th/0701035]; 
W.~Weise, R.~H{\"a}rtle, Nucl. Phys. A {\bf 804}, 173 (2008).

\bibitem{SGM07a} N.V. Shevchenko, A. Gal, J. Mare\v{s}, Phys. Rev. Lett. 
{\bf 98}, 082301 (2007). 

\bibitem{SGM07b} N.V. Shevchenko, A. Gal, J. Mare\v{s}, J. R\'{e}vai, 
Phys. Rev. C {\bf 76}, 044004 (2007). 

\bibitem{ISa07} Y. Ikeda, T. Sato, Phys. Rev. C {\bf 76}, 035203 (2007). 

\bibitem{FGM99} E. Friedman, A. Gal, J. Mare\v{s}, A. Ciepl\'y, Phys. Rev. C 
{\bf 60}, 024314 (1999).

\bibitem{SNR93} M.M. Sharma, M.A. Nagarajan, P. Ring, Phys. Lett. B {\bf 312}, 
377 (1993).

\bibitem{STo94} Y.~Sugahara, H.~Toki, Nucl. Phys. A {\bf 579}, 557 (1994). 

\bibitem{Rij06} Th.A. Rijken, Phys. Rev. C {\bf 73}, 044007 (2006). 
 
\bibitem{TKO07} A.~Martinez Torres, K.P.~Khemchandani, E.~Oset, submitted to 
Eur. Phys. J. A, arXiv:0712.1938 [nucl-th]. 

\bibitem{FKV06} P.~Finelli, N.~Kaiser, D.~Vretenar, W.~Weise, Nucl. Phys. A 
{\bf 770}, 1 (2006). 

\bibitem{GSB98} N.K.~Glendenning, J.~Schaffner-Bielich, Phys. Rev. Lett. 
{\bf 81}, 4564 (1998). 

\bibitem{Ame86} S.R.~Amendolia {\it et al.}, Phys. Lett. B
{\bf 178}, 435 (1986).

\bibitem{SSB80} G.~Simon, C.~Schmitt, F.~Borkowski, V.~Walther, Nucl. Phys. A
{\bf 333}, 381 (1980).

\end{thebibliography}
\end{document}